\documentclass[a4paper,fleqn,usenatbib]{mnras}

\usepackage{newtxtext,newtxmath}
\usepackage[T1]{fontenc}
\usepackage{ae,aecompl}

\usepackage{graphicx}	% Including figure files
\usepackage{amsmath}	% Advanced maths commands
\usepackage{amssymb}	% Extra maths symbols
\usepackage{subfigure}
\usepackage{float}
\usepackage{hyperref}
\title[UV GW Counterparts with GLUV]{Capability of Detecting Ultra-Violet Counterparts of Gravitational Waves with GLUV}

\author[R. Ridden-Harper et al.]{Ryan Ridden-Harper,$^{1}$\thanks{E-mail: ryan.ridden-harper@anu.edu.au}
B. E. Tucker,$^{1,2}$
R. Sharp,$^{1}$
J. Gilbert$^{1}$ 
and M. Petkovic$^{1}$
\\
$^{1}$Research School of Astronomy \& Astrophysics, Mount Stromlo Observatory, The Australian National University,\\ Cotter Road, Weston Creek, ACT 2611, Australia\\
$^{2}$ARC Centre of Excellence for All-Sky Astronomy (CAASTRO)
}

\date{Accepted 2017 August 28. Received 2017 August 21; in original form 2017 June 15}

\pubyear{2017}

\begin{document}
\label{firstpage}
\pagerange{\pageref{firstpage}--\pageref{lastpage}}
\maketitle

\begin{abstract}
With the discovery of gravitational waves (GW), attention has turned towards detecting counterparts to these sources. In discussions on counterpart signatures and multi-messenger follow-up strategies to GW detections, ultra-violet (UV) signatures have largely been neglected, due to UV facilities being limited to SWIFT, which lacks high-cadence UV survey capabilities. In this paper, we examine the UV signatures from merger models for the major GW sources, highlighting the need for further modelling, while presenting requirements and a design for an effective UV survey telescope. Using $u'$-band models as an analogue, we find that a UV survey telescope requires a limiting magnitude of m$_{u'}\rm (AB)\approx 24$ to fully complement the aLIGO range and sky localisation. We show that a network of small, balloon-based UV telescopes with a primary mirror diameter of 30~cm could be capable of covering the aLIGO detection distance from $\sim$60--100\% for BNS events and $\sim$40\% for BHNS events. The sensitivity of UV emission to initial conditions suggests that a UV survey telescope would provide a unique dataset, that can act as an effective diagnostic to discriminate between models.
\end{abstract}


%%%%%%%%%%%%%%%%%%%%%%%%%%%%%%%%%%%%%%%%%%%%%%%%%%

%%%%%%%%%%%%%%%%% BODY OF PAPER %%%%%%%%%%%%%%%%%%

\section{Introduction}

The historic discovery of the first gravitational waves (GWs) from the coalescence of binary black hole systems \citep[BBH;][]{Abbott2016,Abbott2016a}, has drawn attention to a new way of investigating the Universe. The first event GW150914 was inferred to have initial masses of $29^{+4}_{-4}M_\odot$ and $36^{+4}_{-5}M_\odot$, a final merged mass of $62^{+4}_{-4}M_\odot$, and occurred at a luminosity distance of $410^{+160}_{-180}Mpc$ \citep{Abbott2016}. While the second event GW151226 occurred at a luminosity distance of $440^{+180}_{-190}Mpc$, it was a lower mass event, with a final merged mass of $20.8^{+6.1}_{-1.7}M_\odot$ \citep{Abbott2016a}. The third event GW170104 is the most distant event detected, so far, at a luminosity distance of $880^{+450}_{-390}$~Mpc, with a final merged mass of $50.7^{+5.9}_{5.0}$ $\rm M_\odot$  \citep{Abbott2017}.

Although a seminal moment in history, the data obtained from the GW events was limited. Without any complimentary information to the GW detection, the discovery is unable to test underpinning aspects to General Relativity, such as the propagation velocity of gravitational waves \citep{Nishizawa2016}. In an effort to detect the GW progenitor systems, a new era in multi-messenger astronomy is forming to complement GW detectors \citep{Coward2011,Kelley2013,Chu2015,Abbott2016b,Abbott2016c,Evans2016}. Theories regarding counterparts to GW sources have become increasingly important as they inform observation strategies and provide testable predictions. The primary sources of aLIGO GW detections are expected to be binary systems of compact objects such as coalescing binary neutron stars (BNS), coalescing black hole and a neutron star (BHNS), and coalescing binary black holes (BBH).

Currently the UVOT instrument on SWIFT is the only UV telescope that has participated in multi-messenger follow-up observations \citep{Evans2016}. The narrow 17$\times$17 arcmin field of view of SWIFT heavily limits the telescope's effectiveness for both follow-up and serendipitous detections of GW counterparts. Given the present lack of UV survey capability, it is unsurprising that detailed UV models for GW counterparts have not been developed. However, it leaves any future UV survey missions largely uninformed on expected signals, particularly for fast UV transients that GW events may produce. To begin the discussion on UV transients from GW events, we review relevant models, using the $u'$-band as a proxy, due to the limitations of UV models, with which we form a case for a capable and versatile UV survey telescope. One reason short wavelengths are poorly modelled is due to unquantified uncertainties produced by lanthanide absorption in ejecta from mergers involving neutron stars \citep[NSs;][]{Kasen2013,Kasen2015}.

The motivation for this analysis of UV counterparts to GWs stems from a UV survey telescope under development, known as GLUV. Described in \cite{Sharp2016}, GLUV will be a balloon--based near-UV survey telescope, with the primary objective of high-cadence, early UV observations of supernova. During the first hours to days of a supernova, interactions between the ejecta and outer layers or companions are expected to produce UV bright shock emissions \citep{Falk1978,Klein1978,Kasen2010}.

The instrument will be low-cost, high-cadence, and able to provide UV observations from both hemispheres. As this telescope will present a unique opportunity in UV astronomy, we analyse the benefit it would provide in the study and characterisation of GW sources. 

A number of different UV survey missions are currently being considered. As previously mentioned GLUV will be balloon--based, however other proposed systems, such as ULTRASAT intend to be space based, utilising cubesat technology \citep{Sagiv2014}. Being space based, ULTRASAT is not limited by atmospheric transmission, thus, the bandpass is expected to extend into the far-UV (200--240~nm). The proposed wavelength ranges of GLUV and ULTRASAT are complementary, and if both instruments are successful significant survey capabilities would be available for far-UV and near-UV wavelengths. The review and analysis that follows are complimentary to any future UV survey system.

The UV signatures of GW mergers are discussed in Section~\ref{sec:mergers}. In Section~\ref{sec:telescope} we present the preliminary GLUV telescope specifications, survey configurations, expected detection rates and consider the use of GLUV as a complementary data source to the upcoming Large Synoptic Survey Telescope (LSST). LSST will feature a wide field of view, with a baseline cadence of $\sim$3 days to map the Southern sky in optical to near infra-red wavelengths \citep{Ivezic2008}. Since GLUV will be operating at UV wavelengths, a case is made for how its observations would complement those of LSST to characterise GW events.

All magnitudes presented are AB.

\section{UV signatures from mergers}
\label{sec:mergers}
In this section we will explore the luminosity of GW counterparts in the $u'$-band as a proxy for the UV. Although the models presented in the section feature larger uncertainties for shorter wavelengths, they provide a powerful guide to benefits offered by UV observations.

\subsection{Binary black hole merger}
Optical counterparts to BBH mergers are widely unexpected, however, the possible detection of a short gamma ray burst (sGRB) by the Fermi Gamma-ray Burst Monitor 0.4s after GW150914 \citep{Connaughton2016} spurred discussion of multi-messenger counterparts. A comprehensive study of the multi-messenger follow-up survey for GW150914 in \citet{Abbott2016c} found the sGRB was unrelated and found no electromagnetic counterparts. Despite this non-detection, theory developed to support the initial sGRB claim suggests it may be possible to detect an optical and UV counterpart if an accretion disk is present in the BBH system.

The origin of an accretion disk may stem from either the accretion of dense interstellar dust if the BBH formed through direct collapse as discussed in \citet{Belczynski2016}, or via the formation of a ``fossil disk" from fall-back material of a failed supernova, \citep{Perna2014}. A fossil disk is expected to be a product of super-Eddington accretion winds in the fall-back accretion disk. Such a process would produce a bright UV transient, reaching M$_{u'}\rm(AB)=-16$ for a Wolf-Rayet star and M$_{u'}\rm(AB)=-17.5$ for a blue super-giant \citep{Kashiyama2015}. A UV survey telescope could be able to spot the formation of fallback BHs and potentially the systems in which BBH mergers with a fossil disk could occur. 

As described in \citet{Murase2016}, a fossil disk would be long-lived and undetectable until it is disrupted and ionised by the BH-BH merger. The idea of a long-lived fossil disk, has been questioned by \citet{Kimura2017}, who show that the inclusion of tidal torque causes the disk to become excited and active, thousands of years before the BBH merger. The fossil disk hypothesis is likewise examined by 
\citet{Ioka2016} who argue that accretion of the interstellar medium will heat the disk, greatly reducing its lifetime. These two extensions to the fossil disk model highlight the uncertainty in current understand of EM counterparts to BBH mergers. 

In the case of GW150914, \citet{Murase2016} show a fossil disk of ${10^{-5}-1\rm M_\odot}$ is heated and ionised during the merger to a thermal emission temperature of $1.1\times10^4~$K. By integrating the black body distribution over the $u'$-band, it is found that 17\% of the accretion disk emission is within the bandwidth. Using the bolometric luminosity, the accretion disk UV luminosity is calculated to be:
\begin{eqnarray}
L_{UV} & \approx & 6.12\times10^{39} \;\rm erg\;s^{-1}\; \it M_{BH,1.78}\,R^{-\frac{1}{3}}_{d,8} \nonumber \\  && \times\, r^{\frac{1}{3}}_{w,10.5}\, \left(\kappa_T/0.34\; \rm cm^2\; g^{-1}\right)^{-1}, 
\end{eqnarray}
where $\rm M_{BH}$ is the BH mass, $\rm R_d$ is the disk radius, $r_w$ is the radius at which the disk wind is no longer a continuous outflow, $\kappa_T$ is the Thompson scattering opacity and the subscript numbers denote the power of 10 with which the variable is normalised. The luminosity of the accretion disk is linearly dependent on the final merged BH mass, so larger BH mergers will have larger luminosities, thus easier to detect. 

\begin{figure}
	\includegraphics[width=\columnwidth]{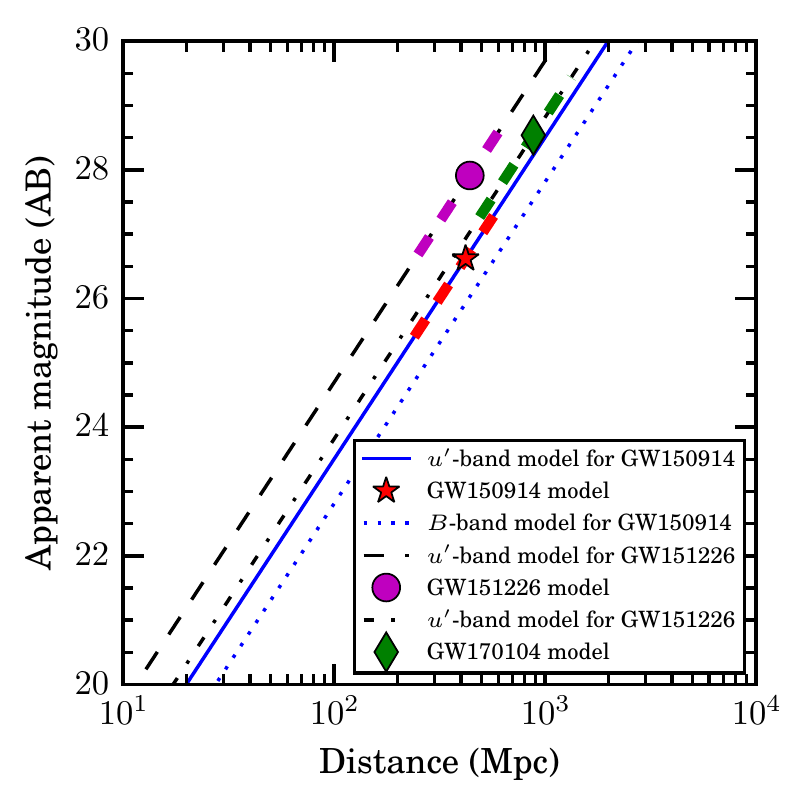}
	\caption{Distance-magnitude relation for both BH-BH merger events detected by aLIGO, calculated using the bolometric luminosity in \citet{Murase2016}. The thick dashed lines indicate the distance uncertainty in the measurements.}
	\label{fig:BHBH}
\end{figure}

The emission models for the first two BBH detections are shown in Fig.~\ref{fig:BHBH}. It is apparent that the disk winds for this model is half a magnitude brighter in $B$-band than $u'$-band, but the disk winds are not the only emission source expected. \citet{Perna2016} and \citet{Murase2016} show that super-Eddington accretion will likely produce a GRB, which is expected to produce a prominent, UV afterglow.

The likelihood of detecting disk wind emission from a BBH merger is small. Detection is challenging due to the short event lifetime of $\sim$3 hours, defined by the time required for the outflow to reach the photospheric radius. The short lifetime is then compounded by three other factors: 1) a faint absolute magnitude, preventing the aLIGO range from being covered; 2) poor event localisation, inhibiting rapid follow-up, which is required for events with short lifetimes; and 3) the unknown rate of a BBH systems forming with a fossil accretion disk. As GW localisation improves, follow-up surveys will be able to effectively cover the search area. However, this will not improve the aforementioned points 1 and 3, which will dominate the detectability of such BBH configurations.

The model developed in \cite{Murase2016} requires, rapid, half-hourly follow-up will be required to detect EM emission from a BBH merger. The disk wind emission model also indicates that large survey telescopes operating in optical wavelengths, such as the LSST, may be better suited for detecting BBH counterparts. However, expanding the fossil disk model to include tidal torques and interstellar medium accretion raises doubt on the fossil disk model, leaving it unknown if BBH mergers produce EM counterparts, thus, observations in all wavelengths are essential to test all possibilities.

\subsection{Binary neutron star merger} \label{sec:BNS}
The merger of a binary NS (BNS) system is predicted to generate a relatively weak GW with an appreciable optical counterpart known as a kilonova \citep{Metzger2014}. The detection of BNS mergers, both through electromagnetic and gravitational radiation, is expected to answer long standing questions regarding the structure and composition of NSs \citep{Takami2014}. Currently the aLIGO off-line analysis  is expected to detect GW from BNS out to a range of $\sim 70$Mpc for a progenitor with a component mass distribution of $\rm 1.35\pm0.13 M_{\odot}$ \citep{TheLIGOScientificCollaboration2016}. The detection range is expected to increase to $200$Mpc in the future \citep{Abadie2010}.

Despite aLIGO not realising its full detection range, BNS mergers are expected to be detected in the upcoming 3rd aLIGO run. If no BNS events are discovered, a number of models, will be at odds with observation \citep{TheLIGOScientificCollaboration2016}. At peak capability, aLIGO is expected to detect on the order of 40 events per year. With the potential of BNS GW events in the near future, multi-messenger follow-up observations will be crucial to understanding the events.

A BNS merger is expected to emit light in three ways over different time frames;
\begin{enumerate}
\item a neutron jet powered precursor
\item accretion disk outflows following the merger
\item sGRB afterglow
\end{enumerate}
as different physical processes produce these three sources, detection would provide valuable insight into the merger dynamics.

\begin{figure}
	\includegraphics[width=\columnwidth]{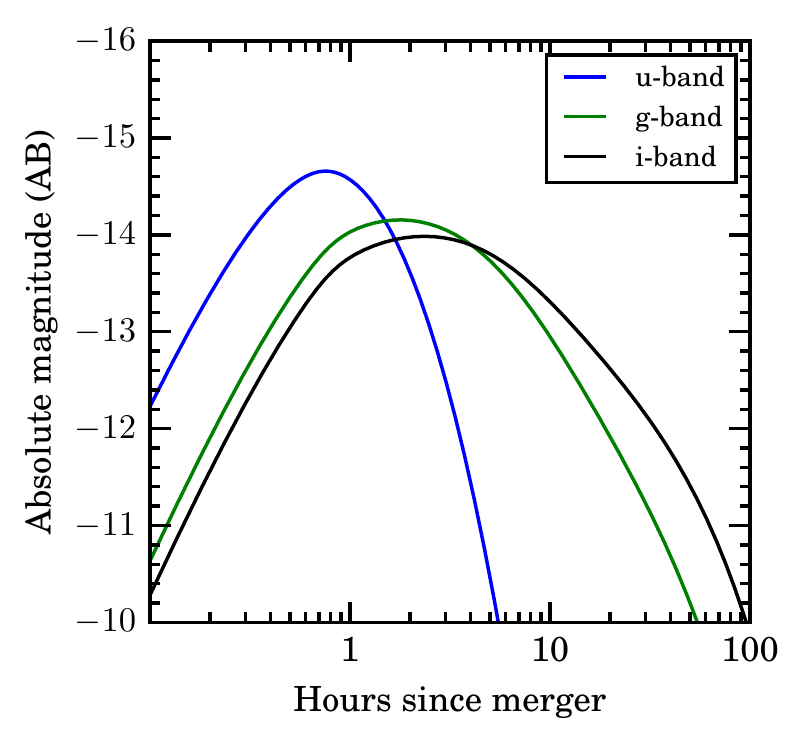}
	\caption{The fiducial neutron-powered precursor, recreated with data provided by \citet{Metzger2015}.}
	\label{fig:Metzger}
\end{figure}

During the merger, it is predicted that at the point of collision jets of mildly relativistic free neutrons could be emitted. Such jets would heat the medium as the neutrons decay, leading to a UV bright precursor \citep{Metzger2015}. The neutron-powered precursor (NPP) is expected have a lifetime of only $\sim$1 hour, as seen in Fig.~\ref{fig:Metzger}.

The primary source of kilonova emission is produced by disk wind outflows from tidally disrupted material \citep{Kasen2015}. As the disk contains a high fraction of neutrons, it is predicted that the r-process will dominate the ejecta increasing the disk opacity at short wavelengths. Since the kilonova luminosity is linked to the disk mass, it depends on both the NS mass ratio and orbital eccentricity \citep{Bauswein2013,Gold2012}.

During a BNS merger, strong magnetic fields on the order of $\sim$$10^{16}$ G can easily be produced \citep{Giacomazzo2014}. Interactions between the magnetic fields and plasma may drive the emission of strong post merger electromagnetic signals and possibly even a sGRB. The nature of sGRBs produced by BNS mergers and their UV afterglow will be discussed in Sec.~\ref{sec:GRB}.

UV emission from BNS mergers are highly model- and remnant- dependent. Short wavelengths offer modelling challenges as the r-process, which takes place in the neutron rich ejecta, can quickly form lanthanide series elements which strongly absorb short wavelengths \citet{Kasen2013}. The r-process can be suppressed by a neutrino flux, which is dependent on the model, thus so too is the spectra. The strong model variation suggests UV observations may be a useful to distinguish between merger pathways.

All models, however, show the same general trend that after $\sim$1 day the kilonova ejecta becomes dominated by the r-process, suppressing all emission at short wavelengths.  Conversely, red wavelengths will benefit from the r-process leading to a long emission fall time.

Unlike BBH or BHNS mergers, BNS mergers have the potential to produce one of four remnants, being: 1) an intermediary high-mass NS (HMNS) that collapses to a BH; 2) a HMNS; 3) a magnetar; and, 4) direct collapse to a  rapidly rotating BH. As each of the remnants will produce different neutrino fluxes, the r-process will proceed at different rates, making UV light curves sensitive to the remnant and its pathway, more so than longer wavelengths.

\citet{Kasen2015} found that if the merger produces a HMNS, the lifetime of the remnant strongly influences the blue emission. For longer lifetimes a HMNS is expected to produce a larger neutrino flux which suppresses the r-process, thus the production of high opacity lanthanide elements is delayed. However, it is also found that a shell of neutron rich ejecta is present around the merger which will act like a lanthanide curtain and heavily reduce UV and blue emission. The variation that is present in the scenarios previously outlined suggest that UV observations would provide powerful diagnostic information on the merger pathway and r-process.
 
If the remnant produced is a magnetar, strong X-ray and UV emission is expected. After the merger, the strong magnetic fields of the magnetar will interact with the surrounding medium and drive an X-ray and UV shock emission which is estimated to be $\sim$100 times brighter than a kilonova and can last upwards of days \citep{Yu2013,Metzger2014}. \citet{Li2016} found that the spin-down of a magnetar could drive out a wind producing an early peak UV magnitude of M$_{u'}\rm(AB)=-21$ a day after the merger. Longer wavelengths have a similar magnitude peak, but days after the merger \citep{Li2016}.
 
To detect a magnetar event within the full aLIGO range of $200$Mpc, a UV survey telescope requires a limiting magnitude of m$_{u'}\rm(AB)\ge 15$. The production of magnetars from BNS mergers would be an excellent candidate for detections in UV, as it should be far brighter in UV than in longer wavelengths.  

\begin{figure*}
	\centering
	\includegraphics[width=\textwidth]{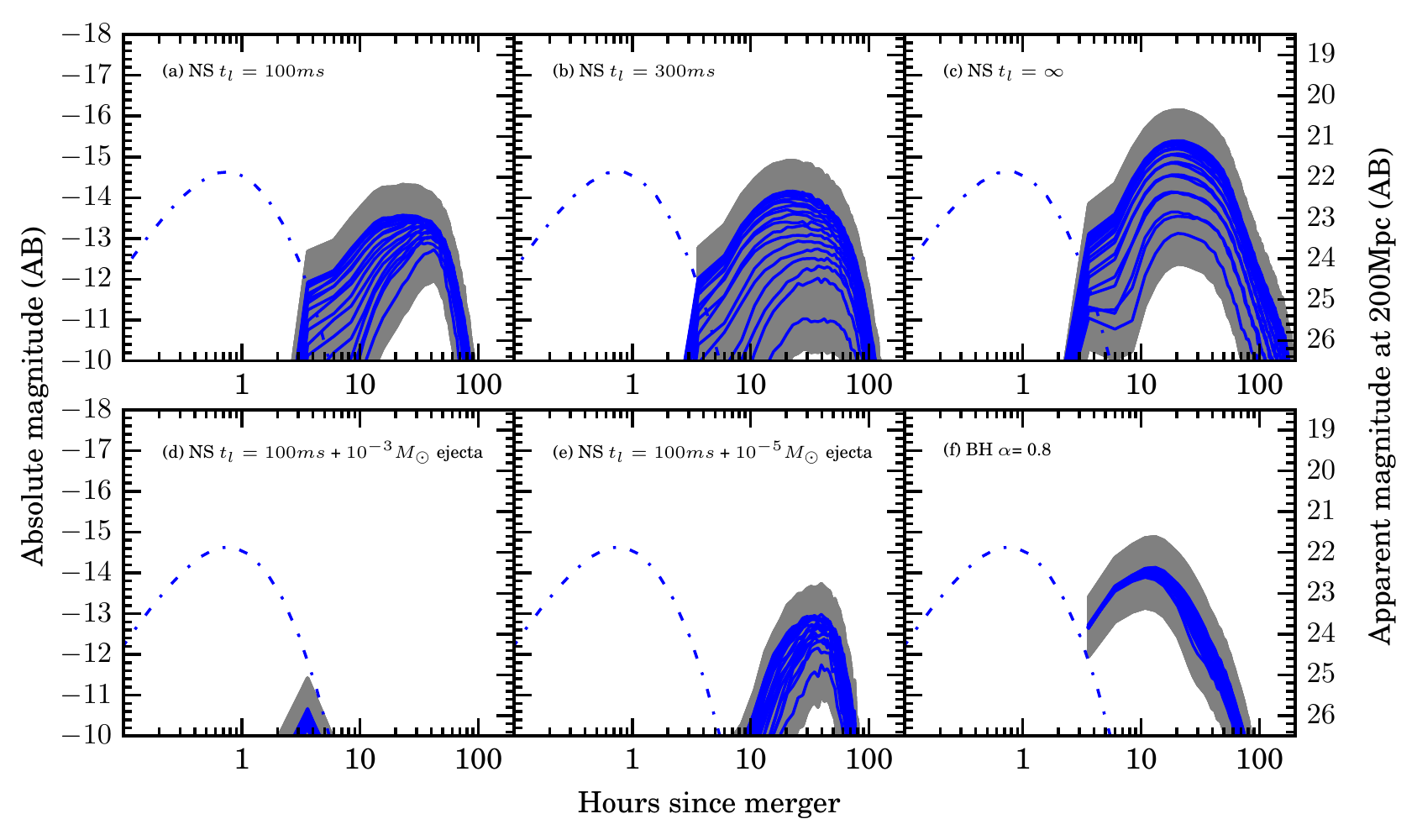}
	\caption{Kilonova $u'$-band light curves for both the neutron-powered precursor (dotted blue) and the disk outflow winds, for viewing angles ranging from $0-180\rm^o$ (solid blue), encased in a factor of 2 error (grey). Figures (a-c) are models which form a HMNS that collapses to a BH after a HMNS lifetime $t_l$. Figures (d,e) are figure (a) with a spherical shell of neutron rich ejecta. Finally figure (f) is a direct collapse to a rapid rotating BH.}
	\label{fig:Kasen}
\end{figure*}

For less exotic remnants, the peak luminosity is expected to be much lower than that of a magnetar. These remnants are expected to have two main sources of UV emission as previously mentioned -- the neutron-powered precursor (NPP) shown in Fig.~\ref{fig:Metzger} and the disk outflows shown in Fig.~\ref{fig:Kasen}.

Further analysis and modelling is required to test the robustness of the NPP since observation angles have yet to be included \citep{Metzger2015,Ferandez2015}. The kilonova disk wind model is also known to underestimate ejecta temperatures, therefore luminosities of shorter wavelengths \citep{Kasen2015}. With the aforementioned sources of error in kilonova emission, the models are taken as the optimal scenario.  

In Fig.~\ref{fig:Kasen} six scenarios developed in \cite{Kasen2015} are shown alongside the NPP. The disk wind model begins at $\sim 2$ hours after the merger, by which time the NPP is expected to have faded. Since the kilonova model doesn't overlap with the NPP, the total early emission from a BNS merger is expected to be larger than the NPP predicts. 

In the cases of an encompassing shell of neutron matter, the NPP offers a baseline of emission as seen in Fig.~\ref{fig:Kasen} (d,e). Although the NPP is short lived, it is expected to be a guaranteed source of UV emission that distinguishes a kilonova produced by a BNS merger rather than a BHNS merger. Thus, early UV observations would be critical to identifying the merger system photometrically \citep{Metzger2015}. 

If a UV survey telescope were to complement the full range of aLIGO (200Mpc), the limiting magnitude must be m$_{u'}\rm(AB)\gtrsim 23$ to detect the neutron-powered precursor. However, for some disk wind models, such as those shown in Fig.~\ref{fig:Kasen} (c) and (f), a limiting magnitude of m$_{u'}\rm(AB)\gtrsim 23$ should be adequate to detect disk outflow winds for most, if not all, viewing angles.

Since the UV emission peaks early, a UV survey telescope would be well suited for rapid follow-up soon after a GW trigger to localise and initiate follow-up observations at longer wavelengths. As the light-curves appear unique at UV wavelengths, early UV observations could be crucial in characterising the merger pathway and understanding the composition of NSs. 

Although all BNS mergers, except those leading to magnetars, will be brighter and long lived at optical wavelengths, the variation and sensitivity of models to UV wavelengths provides an excellent case for discriminating between models and probing elemental abundances. Thus, a UV survey telescope would be able to provide a complementary and perhaps necessary dataset to optical observations.

\subsection{Black hole--neutron star merger}

Another promising candidate for both GW and electromagnetic observations is the merger of a BH and NS system. As there are no known and studied BHNS systems, the detection of GWs or a kilonova associated with such a system would confirm such configurations can occur. The lack of knowledge of BHNS systems gives rise to a rather uncertain rate, however, at full capacity aLIGO is expected to detect BHNS to a distance of $400$Mpc and at a rate of 10 per year \citep{Abadie2010}.

Similar to the BNS case in Sec.~\ref{sec:BNS}, models suggest there may be several processes during the merger that drive EM emission for BHNS. The luminosity of a BHNS merger is dependent on a number of parameters, including the mass and spin of the BH and NS, NS equation of state \citep{Kawaguchi2015}, the NS magnetic field strength \citep{Paschalidis2013,Kiuchi2015,DOrazio2016} and orbital eccentricity \citep{Stephens2011}. 

Shortly before the merger, precursor emission is expected to be generated via interactions between the NS's magnetic field and the BH \citep{Paschalidis2013,DOrazio2016,Paschalidis2015}. Close to the merger, a fireball of $\gamma$ and hard X-rays is expected to be emitted \citep{DOrazio2016}. Such a fireball and gamma-ray burst, discussed further in Sec.~\ref{sec:GRB}, could produce an exciting target for UV survey telescopes.

\begin{figure}
	\includegraphics[width=\columnwidth]{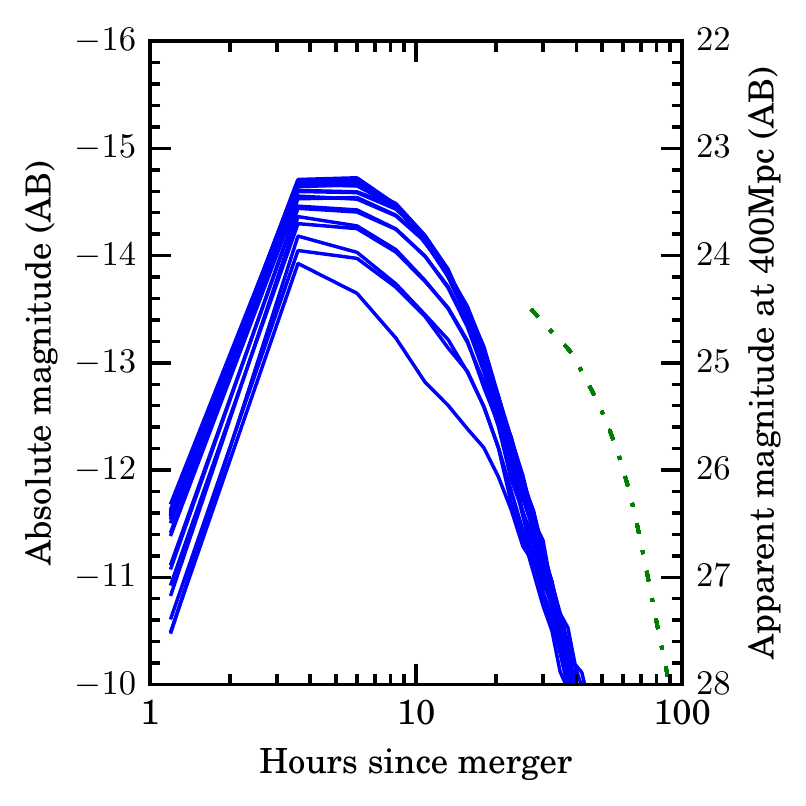}
	\caption{$u'$-band light curves for the two BHNS merger models. In blue the \citet{Fernandez2016} fiducial F0 model, with viewing angles, and in green is the H4Q3a75 BHNS merger model produced in \citet{Tanaka2014} and recreated in \citet{Kawaguchi2016a}. These models exclude early emission sources, such as GRB and fireball afterglows.}
	\label{fig:BHNS}
\end{figure}

BHNS mergers are also expected to produce kilonova from the disk outflows and radioactive decay. As the NS becomes tidally disrupted an accretion disk will form the outflows of which will drive emission similar to that of the BNS kilonova, due to the r-process synthesis of lanthanides \citep{Surman2008,Just2015}. It was found in \citet{Tanaka2014} that BHNS kilonova could, under the right conditions, be brighter than a BNS kilonova. Representative light curves of a BHNS merger from \citet{Fernandez2016} and \cite{Kawaguchi2016a} are shown in Fig.~\ref{fig:BHNS}.   

It can also be seen in Fig.~\ref{fig:BHNS} that the NSBH kilonova light curve is similar that of the BNS kilonova. Both light curves feature an initial peak soon after the merger, that rapidly falls off $\sim$1 day after the merger as r-process elements dominate the ejecta. However, unlike the BNS merger, the BHNS merger is not expected to produce a NPP, so UV observations could be crucial in early classification of kilonova \citep{Metzger2015}.

An afterglow is also possible if some of the NS's magnetic field lines are frozen into the BH. Rotational energy could be extracted from the BH to power a Blandford-Znajek process, this is expected to produce an afterglow, close to the time at which the fireball expands \citep{DOrazio2016}.

As suggested in \citet{DOrazio2016}, if the BH mass is large enough ($\gtrsim 6M_\odot$), it could potentially swallow the NS whole, greatly diminishing or even preventing a kilonova. However, the precursor and afterglow emissions will be unaffected as they are generated by the NSs magnetosphere.

Currently, both the fireball and Blandford-Znajek process afterglows are not modelled in the UV. The lack of afterglow and early kilonova emission heavily limits the ability to asses the ability of a UV survey telescope in BHNS follow-up observations. From Fig.~\ref{fig:BHNS} it can be seen that for a UV survey to completely complement the aLIGO range from the ``late time'' ($>10$ hours) kilonova emission alone, a limiting magnitude of m$_{u'}(AB)\gtrsim 24$ is required. 

Further modelling is required for early time UV emissions, as all BHNS kilonova models, start $\gtrsim 1$ day after the merger. At such late times the UV emission is expected to be small as the r-process is expected to produce a high concentration of lanthanides which will block most UV emission, as with the BNS case \citep{Kasen2013,Tanaka2013}. 
 
\subsection{Gamma-ray bursts}
\label{sec:GRB}
As mentioned in the BNS and BHNS sections, mergers of such objects are expected to produce sGRBs \citep{Tanvir2013,Tanaka2016a,Lazzati2016}. This idea is supported in \citet{Tanvir2013} where a kilonova was linked with GRB130603B, however, it is unclear if the event was produced by a BNS or BHNS merger. The GRBs from these mergers will be highly directional and energetic with afterglows that are visible in the UV \citep{Roming2009}. 

\begin{figure}
	\includegraphics[width=\columnwidth]{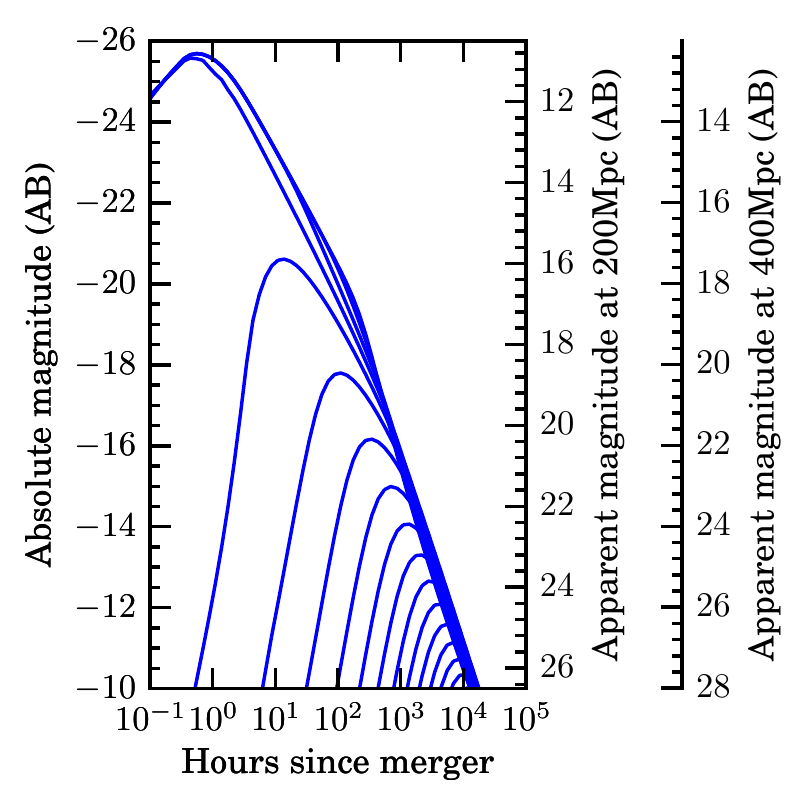}
	\caption{$u'$-band model for a sGRB afterglow produced from a BNS or BHNS merger event, as described in \citet{Lazzati2016}. A range of different viewing angles from 0$\rm ^o$ to 90$\rm^o$ is shown in steps of 5$\rm^o$. Although the sGRB model is capable of producing UV data, $u'$-band data is shown for consistency with the other models.}
	\label{fig:GRB}
\end{figure}

\cite{Lazzati2016} investigate the nature of sGRB afterglows produced by BNS and BHNS mergers. Currently, simulations are not capable of distinguishing a GRB generated by a BNS merger, or a BHNS merger. However they should be different due to variations in mass ratio and cocoon energetics. In Fig.~\ref{fig:GRB} we show the $u'$-band afterglow of a GRB produced from a BNS or BHNS merger at different viewing angles based on models in \citet{Lazzati2016}. 

From the model, it is apparent that a sGRB produced in a merger involving a NS will be bright for many days, even at a distance of 400~Mpc for viewing angles close to the emission axis. Despite the high on-axis luminosity of these events, the magnitude of the afterglow decreases substantially with increasing viewing angles. The strong angle dependence of emission, coupled with the preferential sensitivity of gravitational wave detectors to unaligned systems, joint detections may be unlikely \citep{Lazzati2016}. 

Despite the potential rarity of sGRBs produce by mergers involving NSs, the luminosity of these events make them excellent targets for UV survey missions. For a telescope featuring a limiting magnitude of 22, it would be able to detect an afterglow at  a viewing angle of up to 35$\rm ^o$ at 200~Mpc and 30$\rm ^o$ at 400~Mpc for some days after the merger.

\textit{SWIFT} has been successful in detecting UV light curves from GRBs. Although the UV imager, UVOT, has a small field of view, \textit{SWIFT} utilises the wide field BAT for detection and triggering follow-up observations. This method has lead to a catalogue of UV light curves for GRBs \citep{Roming2009}, however it is unlikely that UVOT could detect afterglows when there is a BAT non-detection. In the case of follow-up observations of an aLIGO GW trigger, it becomes crucial to have UV survey capabilities to detect potential sGRB afterglows, due to the likelihood the SWIFT BAT would be inactive or pointing elsewhere.

\section{GLUV: a UV Survey Telescope}
\label{sec:telescope}
The authors are developing a balloon-based high cadence UV survey to address a key question in supernova physics. During the early stages of a supernova, shock interactions occur which are UV bright and provide crucial information on the progenitor system \citep{Kasen2010}. Operating at UV wavelengths, GLUV is expected to produce routine discoveries of type Ia supernova shock interaction.

As the future of GW science develops, it has become apparent that a UV survey telescope could provide novel data in understanding GW events and their progenitors, using the existing design parameters. Following a brief overview of the instrument, three fiducial observation strategies are presented, along with preliminary rate calculations.

By utilizing advancements in long-duration high altitude balloons, the GLUV project aims to fly a network of UV telescopes at altitudes of 20--30~km for months at a time \citep{Sharp2016}. The telescopes will be modest, with a diameter of 30~cm, recoverable, steerable and feature a 7~deg$^2$ field of view. Although the filter band-pass is yet to be determined, the initial system design employs UV coated CCDs setting the lower limit on the wavelength range of $\sim$250~nm. Further technical details are presented in \citet{Sharp2016}.

Initial evaluations of the system's stability and limiting magnitude appear promising, being $\sim$5~arcsec, m${_{uv} \rm(AB)\approx 22}$ for a signal-to-noise of 5 after a 10 minute exposure, respectively. The bulk characteristics of the system sensitivity are outlined in the following subsections.

\subsection{Sky background}
The faint limiting magnitude is due to the low UV sky background at the expected flight altitude. At expected flight altitude, the effects of scattered light are reduced, resulting in sky glow being the prime contributor to the sky background. The sky background at the intended flight altitudes is currently ambiguous as no current measurements exist. Currently a pathfinder mission is in development to measure the sky background at the intended flight altitudes, the success of which will provide a firm description of the sky background GLUV will experience. 

For a preliminary value of sky background we refer to \citet{Waller1998} which identifies the sky-glow from Oxygen I as m$_{uv} (AB)\approx 26 \rm \;mag\,arcsec^{-2}$. \citet{Waller1998} conducted their measurements from Space Shuttle \textit{Columbia} at a higher altitude than GLUV will fly and at a shorter wavelength of 250~nm, we therefore anticipate GLUV will experience a brighter sky background. For the preliminary calculations, a sky background 2 magnitudes brighter, m$_{uv} (AB)\approx 24 \rm \;mag\,arcsec^{-2}$, is adopted.

\subsection{Instrument throughput}
\citet{Sharp2016} outlines the preliminary design of GLUV. The system will follow a five element catadioptric design containing three corrective lenses alongside the primary and secondary mirrors. The lens elements are expected to have a throughput of 99\%, while the mirror reflectance is taken as 95\%. As a filter is not currently defined, the filter profile is unknown, so a peak throughput of 80\% is implemented. The detector will be a UV coated CCD with a QE~$\approx 55\%$. The resulting instrument throughput is taken to be $\sim$37\%.

Since GLUV is expected to fly close to the ozone layer, atmospheric transmission must be considered. As the flight altitude and therefore atmospheric transmission is yet to be determined, we perform the preliminary calculations assuming the atmospheric transmission to be 40\% at 300~nm. The wavelength limitation by atmospheric transmission is a necessary trade off to ensure that the telescope has maximum flight time.

Future work will identify the true value of atmospheric transmission GLUV will experience. With the inclusion of atmospheric transmission we expect GLUV, with a 30~cm diameter primary mirror, to reach a depth of m$_{uv} \rm(AB)\approx 22$ for 10~minute exposures.

\subsection{Fiducial Survey Strategies \& Detection Rates}

Before considering fiducial survey strategies, it is worth considering how the capabilities of GLUV could complement and compare to upcoming survey telescopes. Two such telescopes are the LSST, an optical survey telescope, and ULTRASAT, a space based UV survey telescope. 

The LSST \citep{LSSTScienceCollaboration2009}, will feature a $\rm9.6\;deg^2$ field of view and work toward a limiting magnitude of $\sim$24.5 for all bands, with a $\sim$3~day cadence. Although LSST is working towards a deep magnitude, the opacity of the atmosphere limits the capabilities in the $u'$-band, with a mean filter efficiency of $\sim$20\%. When taking the atmospheric effects into account, the LSST $u'$-band limiting magnitude is close to what GLUV is expected to achieve. Thus, GLUV would provide an excellent complementary dataset to LSST, due to the sensitivity gain from being at altitude, coupled with a shorter operational wavelength.

ULTRASAT plans to utilise cubesat technology to develop a space-based UV survey telescope. As described in \citet{Sagiv2014}, ULTRASAT is expected to feature a large $\sim$800~deg$^2$ field of view, with a limiting magnitude of $21$ with 12 minute exposures, and have a bandpass of 200-240~nm. In comparison to GLUV, ULTRASAT will have a significantly higher survey rate, however, GLUV is expected to be a magnitude more sensitive, and operate in the near-UV wavelengths ($\sim$250-290~nm). Therefore, these two systems would provide comprehensive, high cadence coverage of the ultraviolet. 

The detection capabilities of GLUV come as a trade-off between cadence and survey area. For the NPP, a cadence of $\sim$30 minutes is required while the kilonova disk winds are longer lived and require $\sim$daily-cadence. Daily-cadence observations limit the maximum survey area to the area one GLUV telescope can observe in a given night. For one GLUV telescope with an on-sky time of 8 hours, the maximum survey area can be calculated as follows;
\begin{eqnarray}
	SA = \frac{8}{\tau}FoV
\end{eqnarray}
where $SA$ is the survey area, $\tau$ is the exposure time in hours, plus 10\% overhead and $FoV$ is the telescope field of view. From signal-to-noise calculations it appears the system will achieve m$_{uv} \rm(AB)\approx 22$ and a signal-to-noise of 5, for a 10~minute exposure. Thus, the maximum survey area for daily-cadence at a depth of m$_{uv} \rm(AB)\approx 22$ would be $\sim300\rm \, deg^2$.

In the event that a constellation of $>$10 GLUVs is flown at any given time, a trade-off can be made between increasing the cadence and increasing survey area. Increasing the cadence, by staggering the constellation longitudinally, would not only provide higher quality light curves, but may also suite the requirements for NPP detection.

To develop a realistic detection rate, the event rates are weighted by an observation probability, constructed as follows. The event rates used in these rate calculations are volumetric, so by using the absolute magnitudes shown in Tab.~\ref{tab:Peakmag} the rates can be converted into a rate per telescope pointing (7 deg$^2$) (see Fig.~\ref{fig:GWrate} for rate comparison). With a rate per pointing, the probability that an event location, $E_L$, occurs within the survey area, $SA$, is given by;
\begin{eqnarray}
	P(E_L\cap SA)=\frac{SA}{\rm Area\;of\;sky}
\end{eqnarray}

We must also account for the probability that an event will be observed in the survey area for a given cadence, $C$, and event lifetime, $E_{lt}$,

\begin{eqnarray}
	P(E_{lt}\cap C)=
	\begin{cases}
		\frac{E_{lt}}{C} &\text{if } \frac{E_{lt}}{C}<1,\\
		1 &\text{if } \frac{E_{lt}}{C} \geq  1.
	\end{cases}
\end{eqnarray}

As each of the conditions stated are independent, the total probability of detecting an event is given by;

\begin{eqnarray}
	P(Detection)= P(E_L\cap SA).P(E_{lt}\cap C)
\end{eqnarray}

With the probability of detection established, the detection rate can be calculated for a number of fiducial survey strategies. 

\subsection{Survey Strategies}

A GW/kilonova survey features two aspects: a follow-up campaign for GW observatories and a ``blind'' transient survey. We will consider both aspects, and their requirements. The survey rate may also be comparable between GLUV and LSST. Although GLUV will be a small instrument requiring 15 minute exposures to reach  m$_{uv}\rm(AB)\approx 22$ for a signal-to-noise of 5, provided the success of initial flights, it is expected that a constellation of GLUVs will be flown. At this time it seems likely that $>$10 GLUVs could fly during a campaign. Individually, the survey area of a GLUV will be far less than LSST, however, if 10 GLUVs are flown they will achieve a survey rate of $\rm 0.080\;deg^2\,s^{-1}$, which is $\sim$12\% that of the expected LSST survey rate. The potential of a large collective survey rate will assist GLUV in providing complementary near-UV observations to LSST.

The follow-up survey requires rapid-response to any GW triggers. From the models presented, it appears that a UV survey telescope with a limiting magnitude m$_{u}\rm(AB)\approx 22$ will be capable of covering 60$-$100\% and $\sim$40\% of the aLIGO detection distance for BNS and BHNS events, respectively. The challenging aspect is likely to come from coordinating observations between a constellation of telescopes. If a GW event is localized to a part of the sky viewable during the night, then a constellation of GLUVs may be adequate for covering the survey area. For localisation on the order of $\sim$400~deg$^2$, a constellation of 10 GLUVs surveying different patches would cover the survey area to a depth of m$_{uv}\rm(AB)\leq 22$ in $\sim$1.4 hours. 

The ``blind'' kilonova survey requires a trade-off between the survey area and cadence. As merger events that produce kilonova are rare, a large survey area is required. Conversely, a primary case for UV observations of kilonova is to detect the NPP, which occurs on short time-scales, requiring high cadence. To address these two aspects we calculate the number of detections per 6 month campaign with a constellation of 10 GLUVs.

\begin{figure*}
	\centering
	\includegraphics[width=\textwidth]{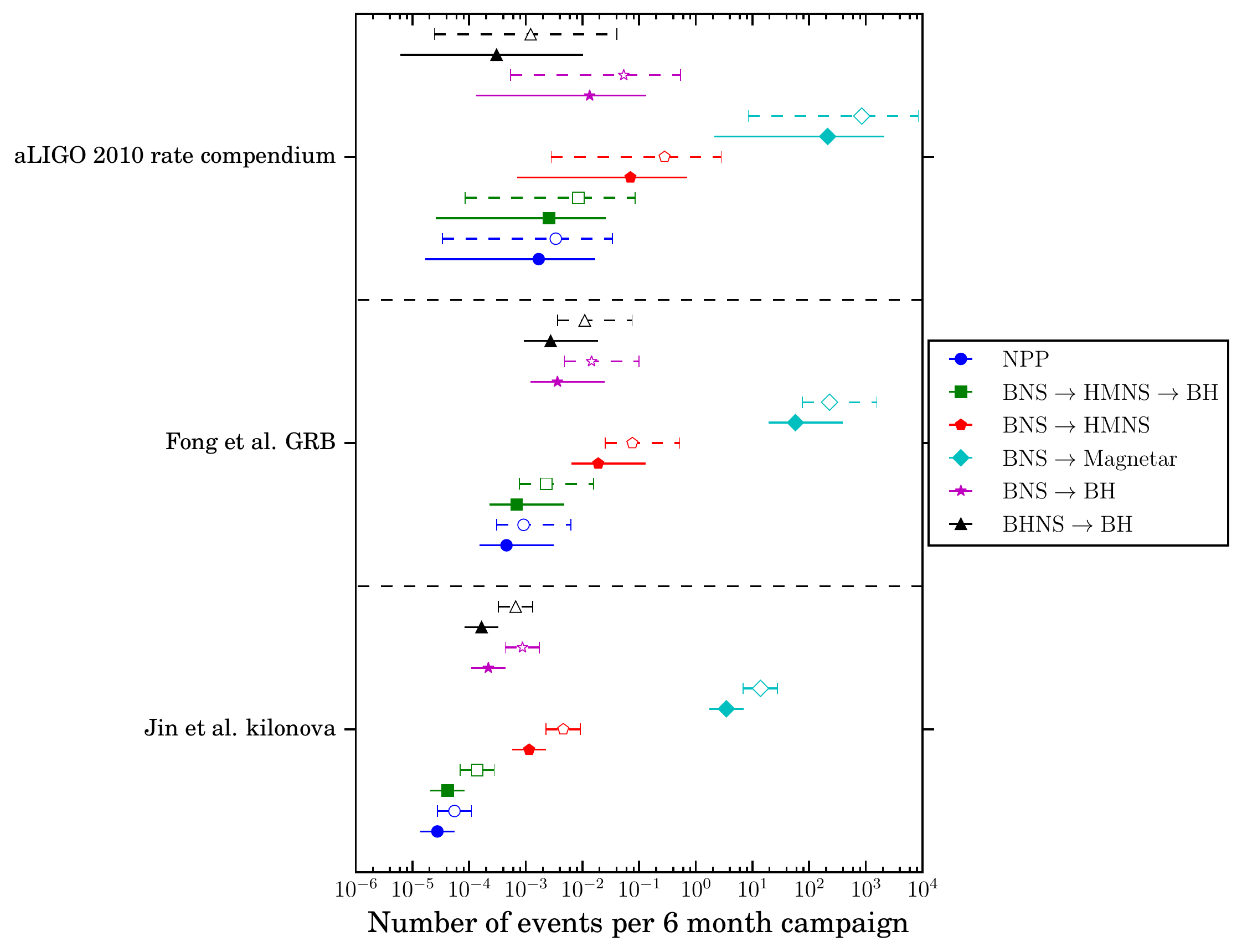}
	\caption{Expected number of GW event detections for a constellation of 10 GLUVs over a 6 month campaign, with limiting magnitude m$_{uv}\rm(AB) = 22$ and two observing strategies. The solid points are for a survey area of 1500~deg$^2$ with 12 hourly cadence, while the hollow points are for a 3000~deg$^2$ with daily-cadence. For these calculations, the full BNS rate is used for each BNS subclass, as it is currently unknown which pathways permitted and at what rates. The  three rates used are, from kilonova \citep{Jin2015}, sGRB \citep{Fong2015} and the aLIGO 2010 rate compendium \citep{Abadie2010}. See the appendix for a breakdown of all rates examined in \citet{TheLIGOScientificCollaboration2016}.}
    \label{fig:GWratesSimple}
\end{figure*}

In the first case the GLUV constellation operates independently, surveying different areas. With each GLUV surveying to a depth of m$_{uv}\rm(AB)\approx 22$, $\sim$3000$\rm\;deg^2$ can be surveyed in a given night. As seen in Fig.~\ref{fig:GWratesSimple}, with the open points, it appears likely that such a survey would detect the luminous UV BNS merger pathways such as $\rm NS\rightarrow HMNS$, $\rm NS\rightarrow Magnetar$ on a regular basis. The $\rm NS\rightarrow BH$ pathway also has a relatively high detection rate which will lead to detections or constrain rates following multiple campaigns.

For the second case, the GLUV constellation monitors 5 independent survey areas with a 12-hour cadence. Again, each GLUV is surveying to a depth of 22. However, they are only covering $\sim$1500$\rm\;deg^2$ which is observed every 12 hours. From looking at the closed points in Fig.~\ref{fig:GWratesSimple}, it is apparent that increasing the cadence by a factor of 2 does not compensate for loss in equivalent survey area.

From preliminary calculations, it appears that survey area is favoured over cadence. A limit of daily-cadence is not required, although, since most events have a lifetime $\sim$1 day, a cadence less than that would likely result in single detections of events. It is also apparent that NPP are unlikely to be observed through a `blind survey' due to their short lifetime. It may be that the best strategy for NPPs is low latency follow-ups of triggers, rather than relying on serendipitous detections.

\begin{table*}
	\centering
	\caption{Peak absolute AB magnitude of GW counterparts, in a range of wavelengths. Cells are left blank if no model data is currently available.}
	\label{tab:Peakmag}
	\begin{tabular}{lccccccr} % four columns, alignment for each
		\hline
		 & NPP & BNS (HMNS$\rightarrow$BH) & BNS (HMNS) & BNS (Magnetar) & BNS (BH) & BHNS & BBH ($\rm 60M_\odot$)\\
		\hline
		GALEX FUV & $-$ & $-$ & $-$ & $-$ & $-$ & $-$ & $-9$\\
		GALEX NUV & $-$ & $-$ & $-$ & $-$ & $-$ & $-$ & $-11.5$\\
 		$u'$-band & $-14.7$ & $-13.4\rightarrow -14$ & $-15.5$ & $-21.2$ & $-14$ & $-14.8$ & $-12$ \\
        $g$-band & $-14$ & $-12.5\rightarrow-15$ & $-15.6$ & $-21$ & $-13.6$ & $-14.6$ & $-12$ \\
        $r$-band & $-13.8$ & $-12.6\rightarrow-15.4$ & $-15.8$ & $-20.9$ & $-13.8$ & $-15.2$ & $-12$ \\
        $J$-band & $-$ & $-14.2\rightarrow -15.6$ & $-16.8$ & $-$ & $-15.4$ & $-16$ & $-11.4$   \\
        
		\hline
	\end{tabular}
\end{table*}

\section{Conclusions}

We have examined the utility of a UV survey telescope for studying gravitational wave sources. Surprisingly, no models of kilonova emission have been developed for short wavelengths including UV. The lack of model data is highlighted in Tab.~\ref{tab:Peakmag}, where entries for NUV and FUV are blank and $u'$-band features unquantified uncertainty.

Although models predict that UV emission will be fainter, than optical for kilonova, it may present valuable diagnostic information to discriminate between models. The sensitivity of UV light curves with respect to models and viewing angles, will be greater than that of $u'$-band, thus, it will provide useful data to test and constrain models, alone and in conjunction with optical wavelengths. Rapid, high energy processes, which are expected to occur in most compact object mergers, will be UV bright. An example is the presence of a neutron-power precursor in a BNS merger, which may be critical in distinguishing kilonova from BNS and BHNS events.

With the limited models available, it appears that a small UV survey telescope could cover a large portion of the predicted aLIGO detection range. Through weighted rate calculations, we find that the detection of BNS mergers may be likely for a constellation of $>$10 GLUVs. Although the kilonova associated with BHNS mergers are expected to be brighter than the BNS counterparts, the low expected rate makes detecting BHNS mergers unlikely.

The fascinating science cases available to UV survey telescopes has prompted the development of such systems. In this paper we have focused on GLUV, which is being developed by the authors with plans to launch in 2019 to conduct a high cadence near-UV survey. Other systems, such as ULTRASAT, aim to provide high cadence UV survey with space based telescopes. If successful the two aforementioned missions would provide a high cadence coverage of a large portion of the UV, opening up new possibilities for studying energetic phenomena and explore short time--scale events. 

We have also compared the effectiveness of GLUV as a complementary near-UV dataset to LSST. Although LSST is expected to feature a faint limiting magnitude, atmospheric transmission renders the limiting magnitude of LSST $u$-band to be close to that of GLUV. Therefore, there is an ideal opportunity for GLUV to provide a complementary dataset at short wavelengths. The survey rate of LSST is far larger than a single GLUV, however, it is expected that in the future, on a $\sim$5 year time--scale, a constellation $>$10 GLUVs will be flying during any given campaign. As the constellation grows, the collective GLUV survey rate may become comparable to that of LSST, while providing a unique, complementary wavelength.

Overall this analysis of UV emission from GW sources has shown a lack of modelling. Without precise models for emissions in NUV wavelengths, we are unable to effectively constrain the utility of a UV survey telescope for detecting GW events. Within the coming years, new UV survey system will be operational, that will open up the possibility of observing and testing models of high energy processes.

\section*{Acknowledgements}
We thank Daniel Kasen for discussions and comments on models cited for both BNS and BHNS mergers, and Davide Lazzati for information on the sGRB afterglow models that were cited. This research was conducted by the Australian Research Council Centre of Excellence for All-sky Astrophysics (CAASTRO), through project number CE110001020 and supported by an Australian Government Research Training Program (RTP) Scholarship.

%%%%%%%%%%%%%%%%%%%%%%%%%%%%%%%%%%%%%%%%%%%%%%%%%%



\bibliographystyle{mnras}
\bibliography{GW.bib} % has a bunch of other unused papers in it, will clean for final version.

%\begin{thebibliography}{99}
%\end{thebibliography}

%%%%%%%%%%%%%%%%%%%%%%%%%%%%%%%%%%%%%%%%%%%%%%%%%%

%%%%%%%%%%%%%%%%% APPENDICES %%%%%%%%%%%%%%%%%%%%%

\appendix

\section{Detection rates}

\begin{figure}
	\centering
	\subfigure[Rates for binary neutron star (BNS) mergers]{\includegraphics[width=\columnwidth]{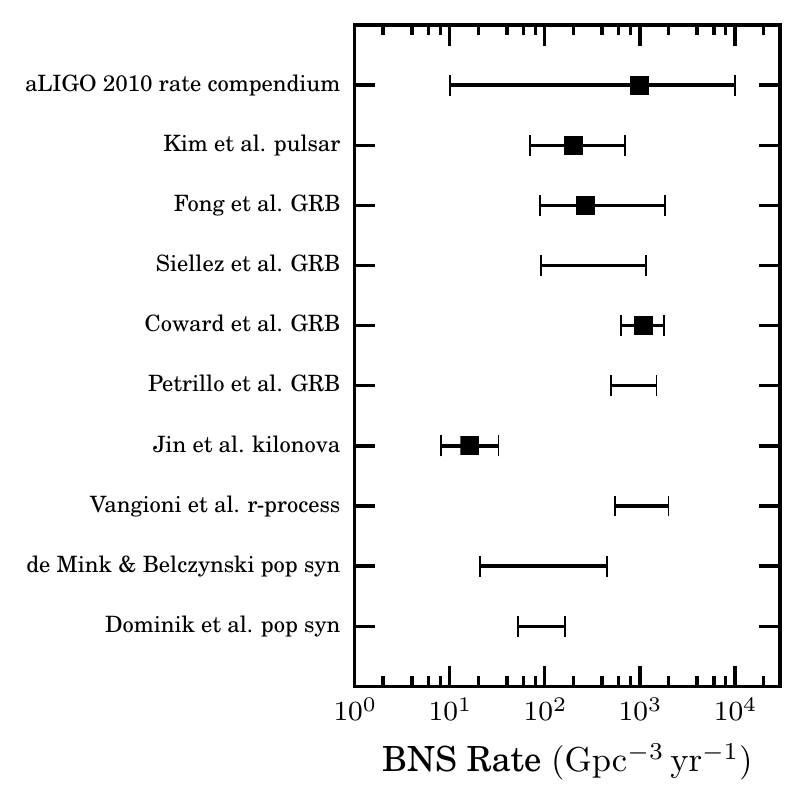}}
	\subfigure[Rates for black hole-neutron star (BHNS) mergers]{\includegraphics[width=\columnwidth]{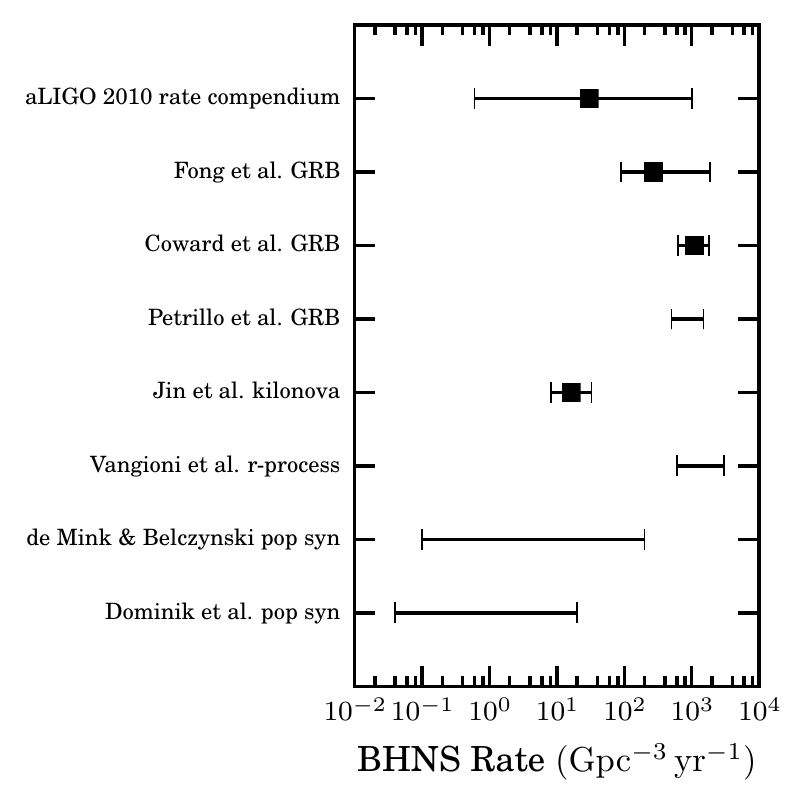}}
	\caption{The merger rates are obtained from a variety of models including: population synthesis \citep{DeMink2015,Dominik2015}, elemental abundance \citep{Vangioni2016}, kilonova rate \citep{Jin2015}, sGRB rate \citep{Petrillo2013,Coward2012,Siellez2013,Fong2015}, pulsar rate \citep{Kim2015} and the aLIGO 2010 rate compendium \citep{Abadie2010}.}
	\label{fig:GWrate}
\end{figure}

Here we present the detection rates for all models examined in \citet{TheLIGOScientificCollaboration2016}. Figures 6 and 7 from \citet{TheLIGOScientificCollaboration2016} have been recreated in Fig.~\ref{fig:GWrate}, to provide a direct comparison between the models.\\

\begin{figure*}
	\centering
	\includegraphics[scale=0.55]{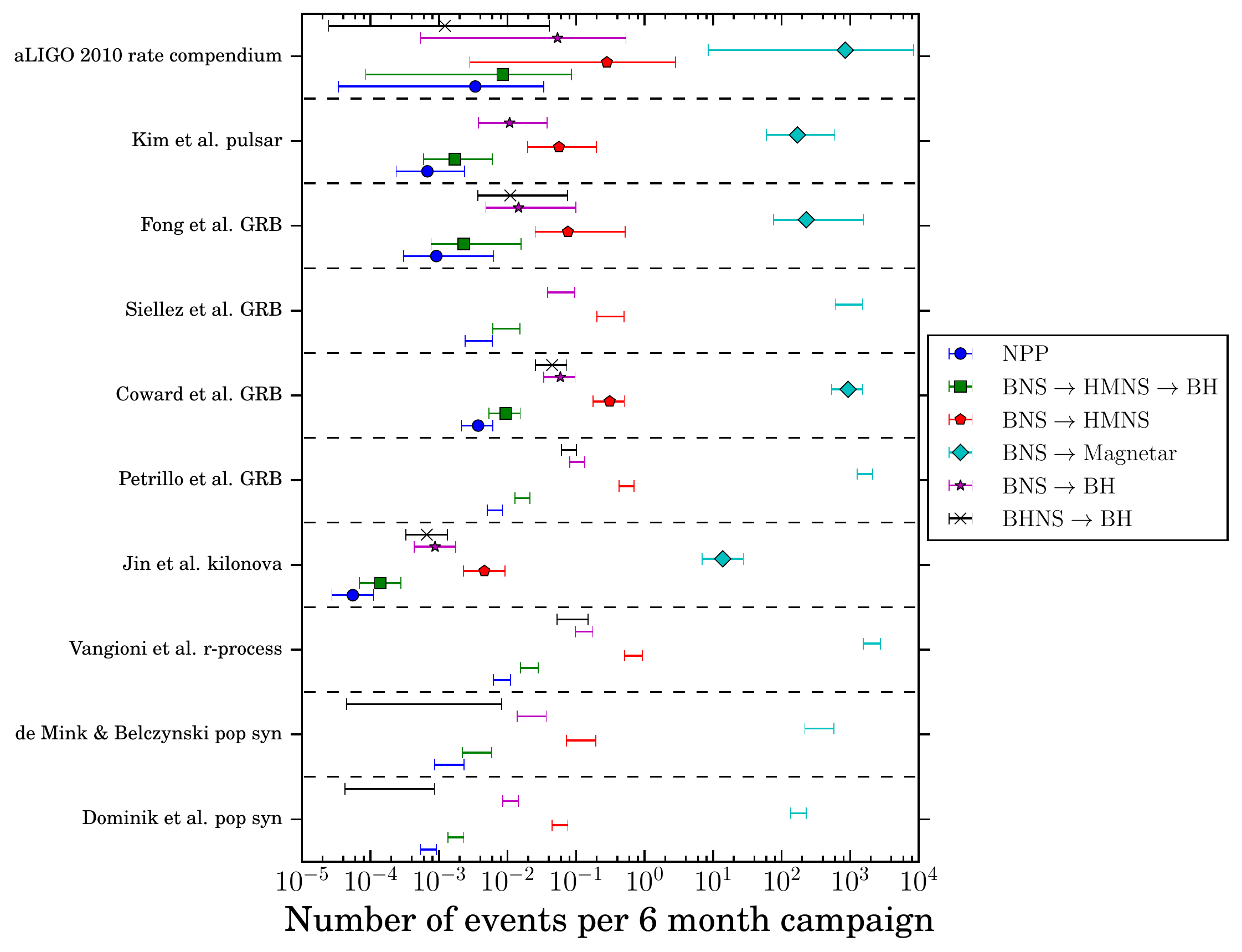}
	\caption{Expected number of GW event detections for a constellation of 10 GLUVs over a 6 month campaign, with limiting magnitude m$_{uv}\rm(AB) = 22$, survey area of 3000deg$^2$, and daily-cadence. The full rate and rate errors for BNS mergers are used for the BNS subclasses, as it is currently unknown which pathways are permitted and at what rates.}
	\label{fig:detections10}
\end{figure*}

\begin{figure*}
	\centering
	\includegraphics[scale=0.55]{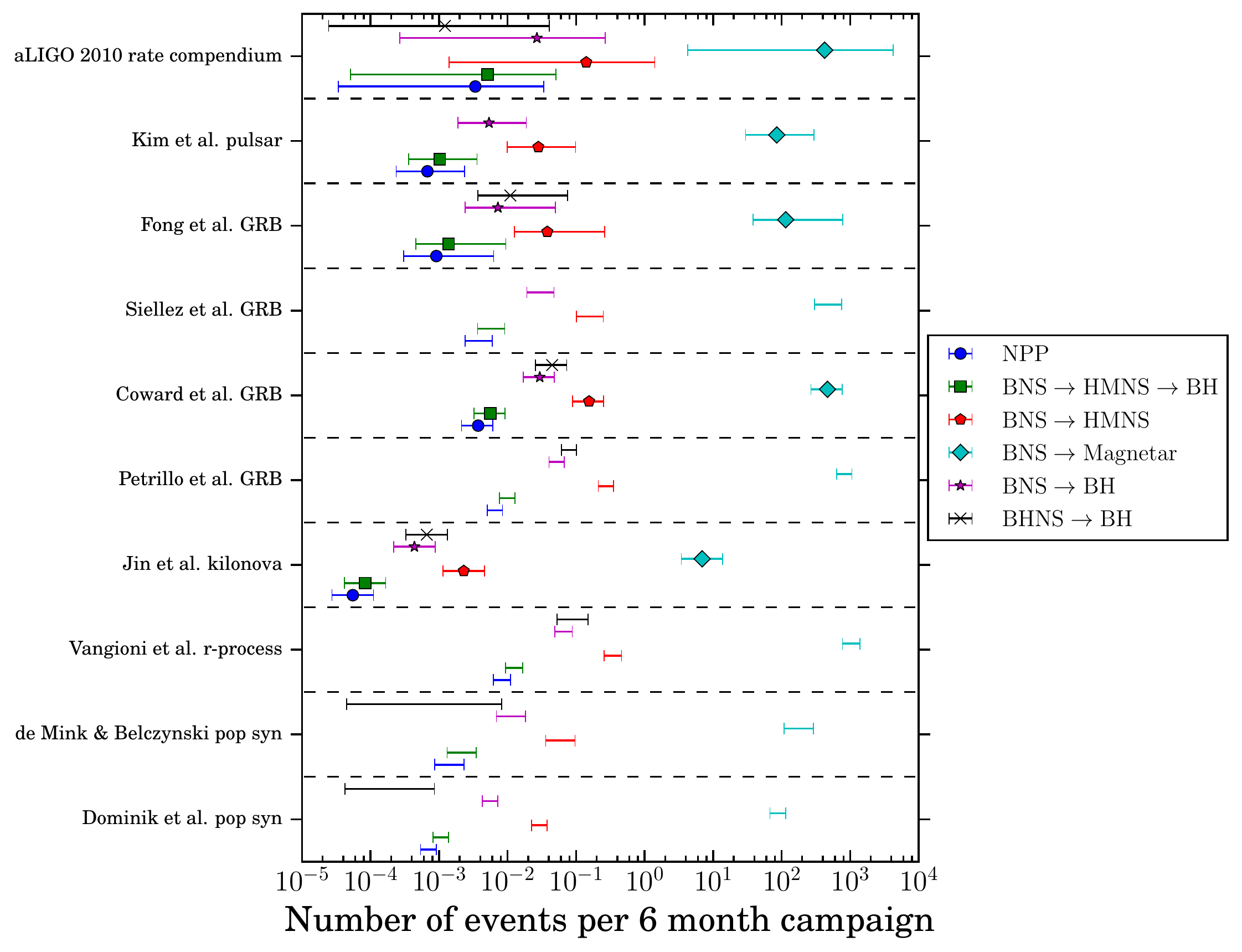}
	\caption{Expected number of GW event detections for a constellation of 10 GLUVs over a 6 month campaign, with limiting magnitude m$_{uv}\rm(AB)= 22$, survey area of 1500deg$^2$, and 12 hourly cadence. The full rate and rate errors for BNS mergers are used for the BNS subclasses, as it is currently unknown which pathways occur and at what rates.}
	\label{fig:detections12}
\end{figure*}

The detections rates for each of the survey configurations are shown in Fig.~\ref{fig:detections10} and Fig.~\ref{fig:detections12}. The spread between the models is encapsulated in the aLIGO 2010 compendium rate.

%%%%%%%%%%%%%%%%%%%%%%%%%%%%%%%%%%%%%%%%%%%%%%%%%%

% Don't change these lines
\bsp	% typesetting comment
\label{lastpage}
\end{document}